\documentclass[aps,twocolumn,showpacs,amsmath,amssymb,pre,superscriptaddress, floatfix,nofootinbib]{revtex4-1} 

\usepackage{amssymb,amsfonts,amsmath}
\usepackage{graphicx}
\usepackage{color}

\newcommand{\1}{\begin{equation}}
\newcommand{\2}{\end{equation}}
\newcommand{\ea}{\begin{eqnarray}} 
\newcommand{\eee}{\end{eqnarray}}
\newcommand{\4}[2]{{\frac{#1}{#2}}}

\begin{document}

\title{Clustering and pattern formation in chemorepulsive active colloids}

\date{\today}

\pacs{05.65.+b,87.17.Jj,82.70Dd,82.40.Ck}

\author{Benno Liebchen}\email[]{Benno.Liebchen@staffmail.ed.ac.uk}\affiliation{SUPA, School of Physics and Astronomy, University of Edinburgh, Edinburgh EH9 3FD, United Kingdom}
\author{Davide Marenduzzo}\email[]{dmarendu@ph.ed.ac.uk}\affiliation{SUPA, School of Physics and Astronomy, University of Edinburgh, Edinburgh EH9 3FD, United Kingdom}
\author{Ignacio Pagonabarraga}\email[]{ipagonabarraga@ub.edu}\affiliation{Departament de F\'{i}sica Fonamental, Universitat de Barcelona-Carrer Mart\'{i} i Franqu\`{e}s 1, 08028-Barcelona, Spain}
\author {Michael E. Cates}\email[]{m.e.cates@damtp.cam.ac.uk}\affiliation{DAMTP, Centre for Mathematical Sciences, University of Cambridge, Cambridge CB3 0WA, United Kingdom}


\begin{abstract}
We demonstrate that migration away from self-produced chemicals (chemorepulsion) generates a generic route to clustering and pattern formation among self-propelled colloids. 
The clustering instability can be caused either by anisotropic chemical production, or by a delayed orientational response to changes of the chemical environment. 
In each case, chemorepulsion creates clusters of a self-limiting area which grows linearly with self-propulsion speed. 
This agrees with recent observations of dynamic clusters in Janus colloids (albeit not yet known to be chemorepulsive). 
More generally, our results could inform design principles for the self-assembly of chemorepulsive synthetic swimmers and/or bacteria into nonequilibrium patterns. 
\end{abstract}

\maketitle

\begin{figure*}
\begin{center}
\includegraphics[width=\textwidth]{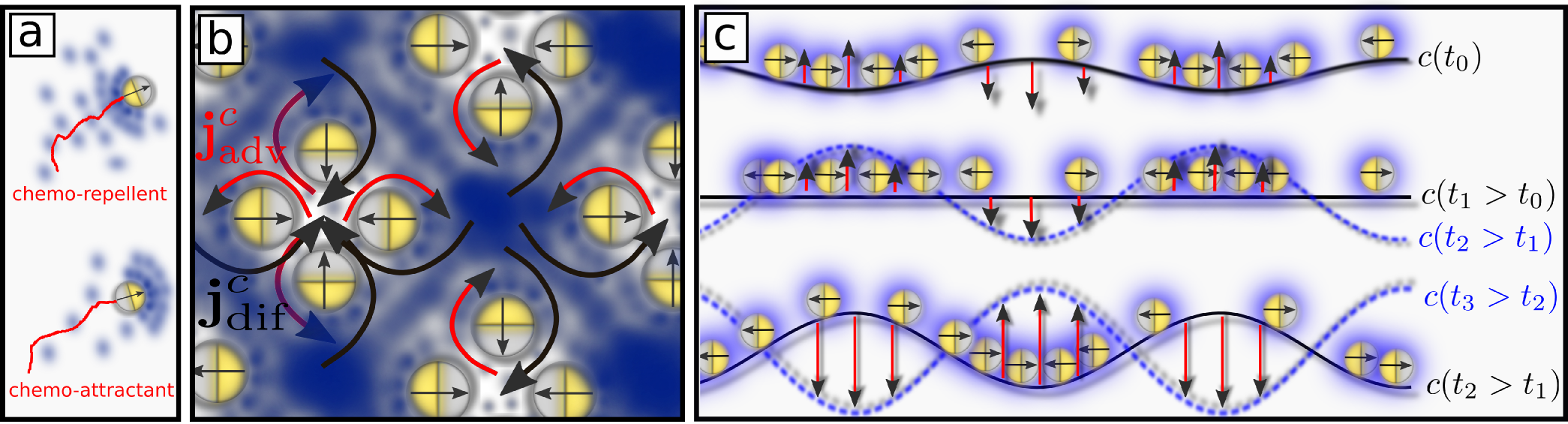} 
\caption{\small Schematics:
a.) Colloidal Janus particles (spheres) half-coated with a catalytic material (yellow) that produces a chemical species (blue); the colloids self-propel by autophoresis in the resulting gradient. 
Chemorepellent colloids propel from high to low concentration of the chemical, chemoattractant colloids vice-versa.
b.) Anisotropy-induced Janus instability, which can lead to aster-like clusters of self-limiting size. 
c.) Delay-induced instability,  
which can lead to travelling wave patterns.
} 
\label{fig1}
\end{center}
\end{figure*}
Active systems, such as suspensions of autophoretic colloidal swimmers, motile bacteria, or other self-propelled particles, are far from equilibrium even in steady state due to their 
continuous energy expenditure~\cite{activematterreview1,activematterreview2}. 
Unlike isothermal Brownian colloids, motile particles can accumulate in regions where they move more slowly. They also slow down at high density, creating a positive feedback loop that can lead ultimately to motility-induced phase separation (MIPS)~\cite{Tailleur2008,Cates2015}.

Experiments with artificial self-propelled colloids have reported self-organised dynamic clustering, sometimes at densities well below that expected to trigger MIPS~\cite{Theurkauff2012,Palacci2013}. These ``living clusters'' seem to reach a limiting size and do not coarsen indefinitely: they show microphase separation rather than macrophase separation as in MIPS. 
So far, the underlying mechanism remains unclear.

In such experiments, motility is autophoretic: a chemical reaction is catalysed on part of the colloid surface, creating a gradient of reagent and/or product that drives the particle forward by diffusiophoresis or a similar mechanism \cite{Golestanian2005,Golestanian2007,Brown2014}. 
The same gradients can then cause chemically mediated long-range interactions between the colloids, and can also cause rotational torques that bias the swimming direction up or down the chemical gradient (an effect known as chemotaxis)~\cite{Saha2014,Hong2007}. 
This fact has suggested a parallel between the experiments in~\cite{Theurkauff2012} and the Keller-Segel (KS) model \cite{Keller1970,Keller1971,Meyer2014} for microorganisms interacting via chemical signalling. This mapping, which assumed that active colloids swim up chemical gradients (the ``chemoattractive'' case), can explain clustering, but leads to complete phase separation, rather than a self-limiting cluster size. A combination of a passive drift up the chemical gradient and self-propulsion down it (``chemorepulsion'') might lead to finite size clusters~\cite{Pohl2014}; however, at variance with experiments~\cite{Theurkauff2012,Buttinoni2013}, these should shrink as the self-propulsion speed increases~\cite{Pohl2014}.
A more general study of chemoresponsive active colloids in the limit of fast chemical dynamics suggests a far larger potential for pattern formation than is predicted by the KS model~\cite{Saha2014}.

Here we propose a theoretical framework for active colloids that describes at a continuum level not only colloidal and chemical densities, but also the local mean orientation (``polarization'') of the active particles. 
We thereby uncover new instability mechanisms which show that chemorepulsion can lead to clustering and pattern formation.
These mechanisms were largely ignored in the literature so far, perhaps because chemoattraction offers a more obvious route to (bulk) phase separation.

We call our two mechanisms the ``Janus instability'' and the ``delay-induced instability''.
The physics underlying these is indicated in Figs.~\ref{fig1}b and c respectively.
For the Janus instability, active particles reorient and move towards a fluctuation-induced minimum in the chemical density, thereby forming an inward-pointing aster (Fig.~\ref{fig1}b). Due to local anisotropy of the chemical production (inevitable for Janus colloids half-coated with catalyst), this cluster produces a shell of chemorepellent, keeping the particles within it together and driving others away, so that it cannot grow beyond a certain size. 
In the delay-induced instability, which appears in a different parameter regime, colloids respond differently to the same fluctuation. Initially they accumulate at the chemical minimum, producing more chemical and lifting the concentration there (Fig.~\ref{fig1}c). 
If they reorient slowly on the timescale needed to cancel the initial fluctuation, the production can overshoot and the minimum becomes a maximum before particles finally drift away. This can trigger a cyclical instability towards traveling wave patterns.
When both instability mechanisms act together, we predict ``blinking'' patterns where clusters continually exchange particles and never reach a steady state.

These chemorepulsive instabilities may shed light on the dynamic clustering or microphase separation recently observed in experiments~\cite{Theurkauff2012,Palacci2013,Buttinoni2013}.
Both instability mechanisms induce clusters of a self-limiting area that grows with self-propulsion speed, which qualitatively agrees with experiments \cite{Theurkauff2012,Buttinoni2013}.
More generally, chemorepulsive instabilities suggest new design principles for the active self-assembly of colloids into spatiotemporal patterns with tunable properties. Our key message that chemorepulsion can generate instabilities of uniform states could also be relevant for biophysics, where the chemoattractive KS instability has long been assumed to drive structure formation among microorganisms~\cite{Gerisch1982,Berg2004,Tindall2008}.

We describe active colloids (living or artificial) at a coarse-grained level, through their density and polarization fields, $\rho({\bf x},t)$ and ${\bf p}({\bf x},t)$. The latter is a local average of the unit vector describing the propulsive direction; this rotates in responds to gradients of a chemical density field $c({\bf x},t)$. 
The colloids self-propel at constant speed $v_0$ and also have isotropic diffusivity $D_\rho$. To represent autophoretic colloids (or signaling bacteria), we assume that the chemical species is produced by the colloids at local rate $k_0\rho$ -- with an important, ${\bf p}$-dependent correction addressed below -- and decays at rate $k_d$. Therefore, we can describe the system dynamics by 
\ea
\dot \rho &=& -\nabla \cdot (\rho v_0 {\bf p}) + D_\rho \nabla^2 \rho \nonumber \\ 
\dot {\bf p} &=& -\gamma {\bf p} + D_{p} \nabla^2 {\bf p} + \beta \nabla c - \gamma_2 |{\bf p}^2| {\bf p} \nonumber \\ 
\dot c &=& D_c\nabla^2 c + k_0 \rho - k_d c + k_a \nabla \cdot (\rho {\bf p}) \label{eom3}
\eee
Here $\beta$ measures the chemotactic coupling strength; when positive, this represents chemoattraction (Fig.~\ref{fig1}a), for instance bacteria swimming up food or aspartate gradients~\cite{Murray2003,Berg2004}. 
Here however we focus on negative $\beta$, describing chemorepulsion, as arises for at least some types of colloid~\cite{Saha2014}, or for cells fleeing from toxins~\cite{Berg2004}. 
In (\ref{eom3}) the polarization decays locally at a relaxation rate $\gamma$, set by rotational diffusion; it also has translational diffusivity $D_{p}\sim D_\rho \sim v_0^2/\tau$~\cite{Cates2015}, which smears out details of ${\bf p}$ on scales below the ``run length'' $v_0/\gamma$.  The term in $\gamma_2$ describes saturation in  ${\bf p}$ at strong alignment \cite{footnote1}.
Finally, $k_a \nabla \cdot (\rho {\bf p})$, where $k_a$ has the dimensions of speed, describes an anisotropic correction to the isotropic chemical production term ($k_0 \rho$), arising whenever the chemical is produced by the colloid asymmetrically \cite{footnote1b,footnote2}.

Rewriting (\ref{eom3}) for dimensionless quantities $\tilde t=k_d t$, $\tilde x= x\sqrt{k_d /D_\rho}$, and setting $\tilde \rho = \rho k_0 |\beta| v_0/(k^2_d D_\rho)$,$\tilde {\bf p}=v_0/(\sqrt{k_d D_\rho}){\bf p}$;$\tilde c=v_0|\beta|/(k_d D_\rho)$ we show (in SI) that the parameter space is spanned by five dimensionless variables $\Gamma=\gamma/k_d;\; \mathcal{D}_p =D_{p}/D_\rho;\; \mathcal{D}_c=D_c/D_\rho; \; \kappa=k_a k_d/(k_0 v_0); \; \Gamma_2=\gamma_2 D_\rho/v_0^2$. Also $\; s={\rm sign}(\beta) \label{ru}$ distinguishing positive ($s=1$) and negative ($s=-1$) chemotaxis. In the following, we omit tildes.

\begin{figure*}
\includegraphics[width=\textwidth]{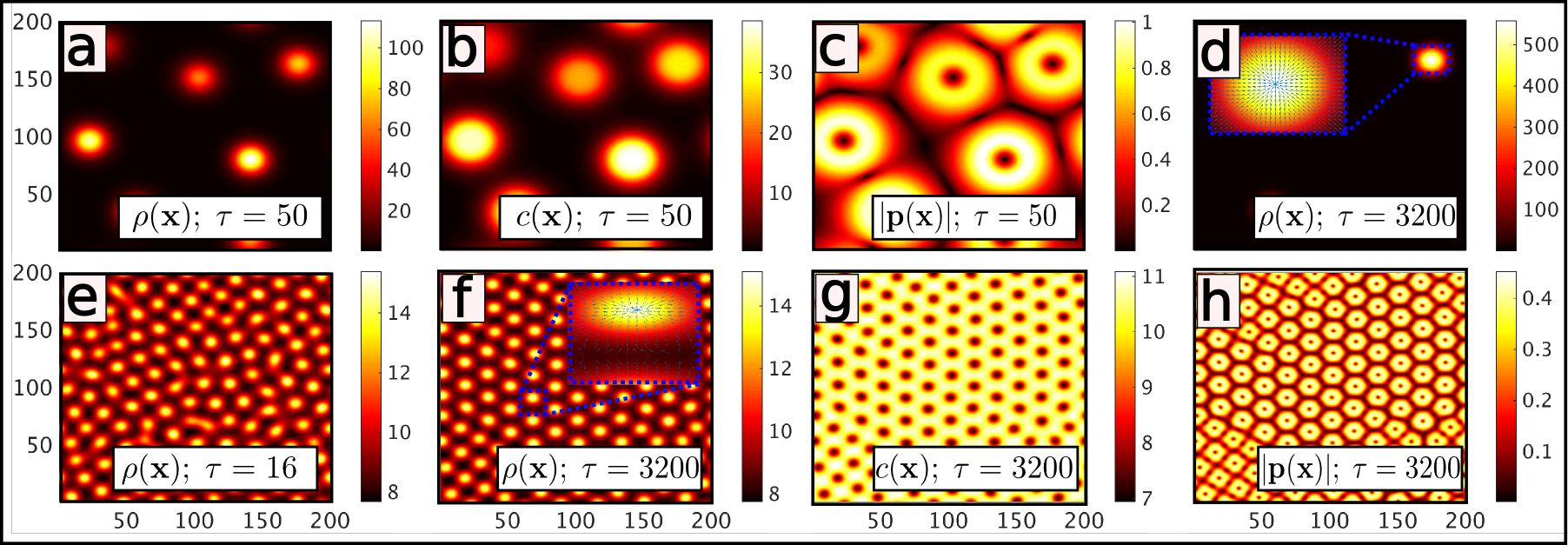} 
\caption{\small Time evolution of the density fields $\rho,c,|{\bf p}|$ (insets: ${\bf p}/|{\bf p}|$) for a weakly perturbed uniform initial state (color shows value of the fields).
(a-d): Clustering and phase separation for chemoattractive colloids ($s=1,\rho_0=8$). (e-h): Arrested phase separation and stationary density pattern for chemorepulsive colloids resulting from the Janus instability ($s=-1, \rho_0=10$).
Other parameters: $\mathcal{D}_p=\mathcal{D}_c=\kappa=\Gamma=1$, $\Gamma_2=10$ and $x_u=5; t_u=1$. (Time and space units of $\tau=t/t_u$ and $x=x/x_u$).}
\label{fig2}
\end{figure*}

We have solved  (\ref{eom3}) numerically on a square box of side $L$ by finite difference methods. Unless otherwise stated, results are obtained for periodic boundary conditions, with as initial condition a small $\rho$-perturbation of the uniform state $(\rho,{\bf p},c)=(\rho_0,{\bf 0},\rho_0)$.
With chemoattraction ($s=1$), the initial uniform state is stable for small $\rho_0$, whereas for stronger coupling it is unstable to the formation of dense colloidal clusters that co-localise with maxima in $c$ (Fig.~\ref{fig2}a,b). 
These droplets coarsen continuously to yield complete phase separation at late times, albeit featuring orientational order in the form of a macro-aster (Fig.~\ref{fig2}d, inset). 
This chemically-induced phase separation is well understood~\cite{Meyer2014}: colloids swim towards high chemical concentration forming a cluster, which increases chemical production locally, recruiting further particles, etc..

Although this feedback loop is absent for chemorepulsion ($s=-1$), upon tuning $\rho_0$ beyond a certain threshold, we observe, strikingly, that the initial fluctuations amplify also for chemorepulsive colloids. 
In marked contrast with the chemoattractive case, the resulting dense colloidal clusters do not coarsen beyond a characteristic size, instead yielding microphase separation (Fig.~\ref{fig2}e-h and Supporting Information (SI), Videos 1,4,5). The behavior en route to this steady state is complex; it can feature amplitude oscillations, or ``blinking" (see SI, Videos 1,4,5). 
Blinking clusters dynamically exchange particles, before settling down into a stationary arrangement, which normally consists of a hexagonal lattice of droplets and an inverted pattern in the chemical density (Fig.~\ref{fig2}g). 
Deviations from the ideal hexagonal structure (visible close to the boundaries in Fig.~\ref{fig2}) can be more or less pronounced, depending on the 
specific parameter choices. As can be seen in Fig.~\ref{fig2}f (inset), chemorepulsive colloids point towards the chemical density minima between the colloidal clusters, and the overall orientational pattern consists of stable asters and anti-asters. Parameter fine tuning, or choosing no-flux boundary conditions instead of periodic ones, can lead instead to permanent blinking (SI, Videos 4,5).

\begin{figure*}
\includegraphics[width=\textwidth]{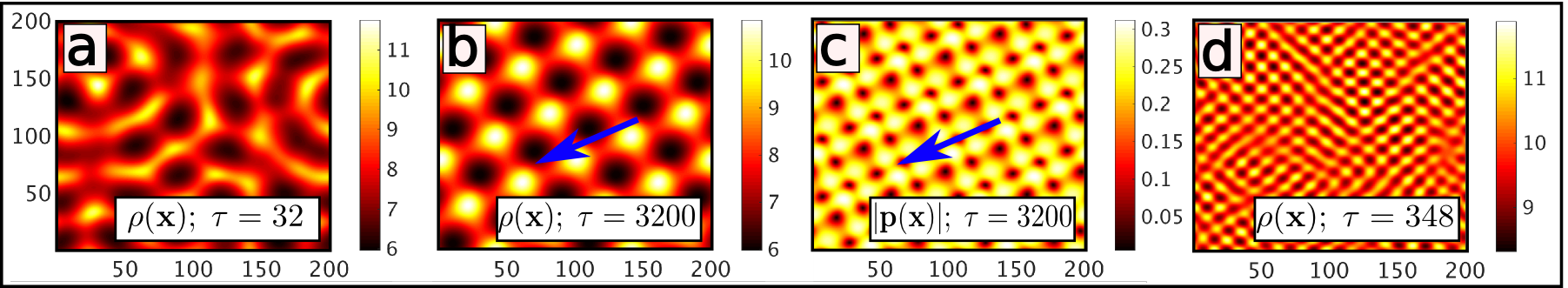} 
\caption{\small 
As Fig.~\ref{fig2}, for $s=-1, \rho_0=8$.
The snapshots show the amorphous transient dynamics (a), travelling wave patterns (b,c) and continously moving states (d)
for chemorepulsive colloids resulting from the delay-induced instability.
Other parameters as in Fig.~\ref{fig2} for (a-c) and $\mathcal{D}_c=0.2; \mathcal{D}_p=1; \kappa=0.5; \Gamma=1.0; \Gamma_2=6.0$ and $x_u\approx 1.83; t_u=1$ for (d).}
\label{fig3}
\end{figure*}

Intriguingly, we also find colloidal waves and travelling oscillatory patterns. In particular, upon further increasing the overall density $\rho_0$ we observe an amorphous pattern which evolves towards a more regular state, where clusters continuously merge, split and decay (Fig.~\ref{fig3} and SI, Video 2,3). 
Eventually, the colloidal and chemical density fields may approach regular lattice-like patterns, which phase-lock and travel at constant velocity along a common direction. This traveling wave is associated with a rectangular lattice, rather than the hexagonal one associated with the stationary patterns (Fig.~\ref{fig2}g). 
The selection of a simple traveling wave is favored by periodic boundary conditions and small system sizes (SI, Video 2); when choosing large systems (SI, Video 3), or imposing no-flux boundary conditions (which might better represent experiments) the pattern persists as a flowing state, continuously forming clusters of well defined size and amplitude (SI, Video 4). 

To further understand the mechanisms of chemorepulsive pattern formation, we have performed a linear stability analysis of our model. 
Assuming that ${\bf p}$ relaxes fast compared to $\rho$ and $c$ and $\Gamma_2 \ll 1$ and ${\cal D}_p \ll 1$ we find ${\bf p}=(s/\Gamma) \nabla c$ (see SI)
and obtain, after rescaling variables $\rho'\equiv \rho/\Gamma;\; c'\equiv c/\Gamma;\; \kappa'\equiv \kappa \Gamma$, a generalized KS (gKS) model~\cite{Murray2003,Meyer2014,Keller1970} that accounts both for chemorepulsion ($s=-1$), and anisotropic chemical production ($\kappa\ne 0$):
\ea
\dot \rho &=& -s \nabla \cdot (\rho \nabla c) + \nabla^2 \rho \nonumber \\ 
\dot c &=& s \kappa \nabla \cdot (\rho \nabla c) + \mathcal{D}_c \nabla^2 c + \rho -c \label{eom3r}.
\eee
For $s = 1$, linear stability analysis (see SI) reveals the standard chemoattractive KS instability when  $\rho_0>1$ (or in physical units, when $\rho_0>\4{D_{\rho}k_d\gamma}{k_0\beta v_0}$)~\cite{Murray2003,Meyer2014}.
For chemorepulsion ($s=-1$) we find a qualitatively different picture, in which instability arises when $\kappa \rho_0>\mathcal{D}_c$, or, in physical units, when $k_a \beta \rho_0>\gamma D_c$. In contrast to the attractive case, the instability is now determined by the anisotropic reaction term; it disappears for isotropic active colloids, $k_a\to 0$. Hence we refer to this as the ``Janus instability''. 
In marked contrast to the classical KS case for chemoattraction, this is a short wavelength instability, arising only for $q>q^*=\sqrt{(1+\rho_0)/(\kappa \rho_0 - \mathcal{D}_c)}$ (Fig.~1 in SI). For large $\rho_0$, or large $\kappa$, the corresponding length scale is $l^*\sim {\kappa}^{1/2}$ ($l^*\sim \sqrt{(D_{\rho}k_a)/(k_0v_0)}$ in physical units), and diverges when the isotropic production rate vanishes. 
Remarkably, the steady-state cluster domain area grows as ${l^\ast}^2 \propto v_0$ (assuming $D_\rho\propto v_0^2$\cite{Cates2015,Theurkauff2012}), a prediction which turns out to remain valid for Eqs.~(\ref{eom3}) even with finite values of $D_p \propto v_0^2$. This may explain recent observations of self-limiting active clusters, whose particle number increases linearly with $v_0$~\cite{Theurkauff2012,Buttinoni2013}. 

Why does the Janus instability lead to arrested, not full, phase separation? We find that two effects limit cluster growth. 
First, each cluster creates a shell of chemorepellent which drives away colloids passing nearby, hindering their arrival. 
Second, as the cluster increases in size, so does the quantity of chemorepellent created at its core via the isotropic production term: once too large, this disintegrates the cluster. 

Note that the gKS model predicts a short wavelength divergence of the growth rate, ${\rm Re}[\lambda({\bf q})]$ as $|{\bf q}|\rightarrow \infty$, suggesting the growth of point-like clusters. We show in SI that rotational dynamics destroys this high $q$-divergence. Thus rotation provides an essential ingredient for chemorepulsive pattern formation, not fully captured by (\ref{eom3r}).

\begin{figure}
\begin{center}
\includegraphics[width=0.48\textwidth]{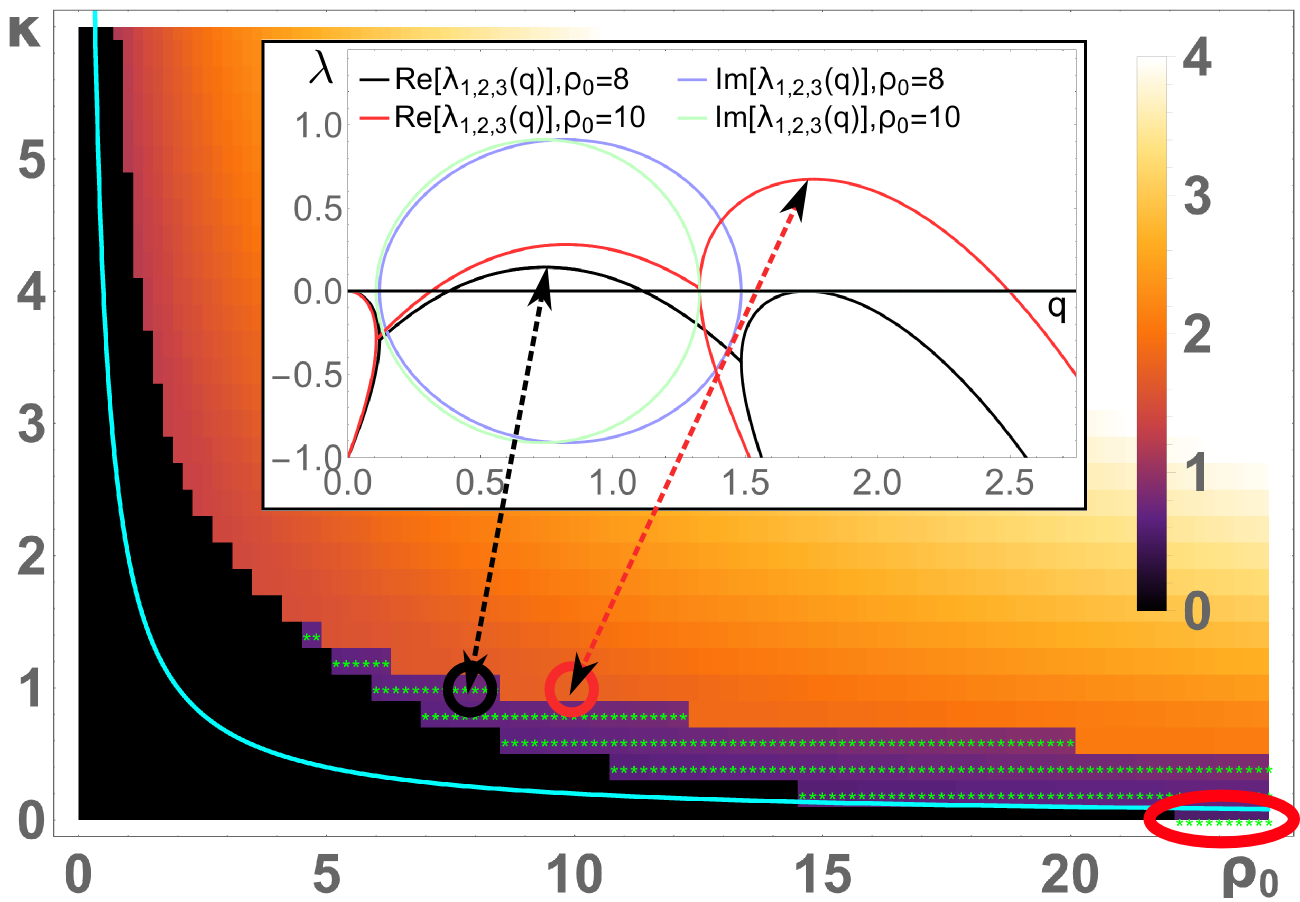} 
\caption{\small Phase diagram showing instabilities of the uniform state for Eqs.~(\ref{eom3r}) 
and $\mathcal{D}_c=\mathcal{D}_p=\Gamma=1$. Black denotes stability; 
colors show the wavelength of the fastest growing unstable mode and green stars show where this mode is oscillatory. 
The red ellipse highlights instability for $\kappa=0$.
The cyan line shows the onset of instability as predicted by the gKS model (\ref{eom3r}). 
Circles mark the parameter sets of Fig.~\ref{fig2}e-h and Fig.~\ref{fig3}a-c.
Inset: Dispersion relation for the encircled parameters (a third solutions, not shown, is real and negative). Upper arrow heads: fastest growing mode.} 
\label{fig4}
\end{center}
\end{figure}

A general, numerical evaluation of the dispersion relations $\lambda(q)_{1,2,3}$ for the three distinct modes that emerge from the full model (\ref{eom3}) allows us to plot phase diagrams on the $\rho_0,\kappa$ plane; a typical example is shown in Fig.~\ref{fig4}. In general, such phase diagrams show three different regimes: (i) one in which the uniform state is stable (black in Fig.~\ref{fig4}), (ii) one in which the growth rate of the instability is real and positive, which corresponds to the Janus instability (orange to white); and (iii) one in which there is a non-zero imaginary part in the growth rate  (purple with green stars). 
Modulo small corrections from finite-size effects, the length-scales set by the wave-vectors of maximal growth rate closely match those observed in the simulations of Figs.~\ref{fig2} and Fig.~\ref{fig3}.

Remarkably, the phase diagram shows that the oscillatory instability can develop even for isotropic chemical production, $\kappa=0$, at large colloidal densities (red ellipse in Fig.~\ref{fig4}). 
Accordingly we must have a second instability, distinct from the Janus mechanism which requires finite $\kappa$. (The cyan line in Fig.~\ref{fig4} does not reach $\kappa=0$ at finite $\rho_0$.) 
This can be traced to the finite relaxation rate of ${\bf p}$ which we neglected when deriving the gKS model (\ref{eom3r}). Because the oscillatory unstable mode requires in effect a delayed reorientation of ${\bf p}$, we name this the ``delay-induced instability''. 
Intuitively, it can be understood as follows (see the simplified, 1D cartoon in Fig.~\ref{fig1}c). Chemorepulsive colloids move towards the minima of an initial fluctuation in chemical density (upper panel, solid line), accumulate there, and produce chemicals opposing the original fluctuation. 
Due to a finite response time this production does not stop at uniform density (middle panel) but overshoots, leading to reversal and possible amplification of the initial fluctuation (dashed blue line). 
This cycle repeats (lower panel) and represents a delay-induced feedback loop: an initial fluctuation of the chemical density field triggers another fluctuation of the same field but with opposite sign. 

To understand the delay-induced instability quantitatively, we develop a minimal model in the SI, showing that consecutive fluctuations can amplify if $\rho_0>1$ leading to an oscillatory instability.
Deep in the pattern forming regime ($\rho_0\gg 1$) we find that the wavelength of the fastest growing mode scales as $\rho_0^{1/2}$, or in physical units as $l^\ast \propto \sqrt{|\beta|v_0 k_0\rho_0/k_d^3}$ predicting (in accordance to our numerical simulations) that the cluster area grows linearly with the self-propulsion velocity.
\\The transient or permanent ``blinking" of clusters which we observed in Fig.~\ref{fig2}e appears close to the transition line between the Janus instability and the delay-induced instability (regimes (ii) and (iii) in Fig.~1 of the SI). Here, stationary and oscillatory modes of different wavelength grow with similar rate out of the uniform state and lead to an effective particle motion on top of a stationary density profile which causes the blinking. (See SI for details).

In conclusion, our two chemorepulsive instabilities create a robust new route to pattern formation. Both instabilities lead to clusters of self-limiting area which grows linearly with propulsion speed $v_0$. This agrees with recent experimental observations~\cite{Theurkauff2012,Palacci2013,Buttinoni2013} and may shed light on the still mysterious mechanism underlying their appearance. (Competing explanations based on the chemoattractive KS instability either predict macrophase separation or clusters shrinking with increasing ${v_0}$ \cite{Pohl2014}.)
More generally, these chemorepulsive instabilities might inform design principles for creating active colloids that can self-assemble into spatiotemporal patterns with
desired properties. 
Finally, or key finding that chemorepulsion can generate instability of uniform states might also be important for biophysics, where the chemoattractive KS instability has long been invoked to explain patterns of microorganisms~\cite{Gerisch1982,Berg2004,Tindall2008}. In growing biofilms, for example, the interaction of bacteria with self-secreted polymer~\cite{Ghosh2015} might be interpreted as chemorepulsion.

We thank EPSRC EP/J007404 for funding. 
B.L. gratefully acknowledges funding by a Marie Curie Intra European Fellowship (G.A. no 654908) within Horizon 2020. 
M.E.C. is funded by a Royal Society Research Professorship. I.P. acknowledges the Direcci\'on General de Investigaci\'on
(Spain) and DURSI for financial support under Projects No. FIS 2011-22603 and No. 2009SGR-634, respectively, and Generalitat de Catalunya under program Icrea Acad\`emia.


\begin{thebibliography}{10}
\bibitem{footnote1}
The patterning scenarios we report do not change qualitatively if this cubic saturation term is omitted and a nonlinear chemotactic coupling is instead used, 
e.g. $\beta \nabla c/|\nabla c|$, 
which prevents unlimited growth of $\nabla c$. We favor the cubic saturation term because then $\gamma_2$ does not enter the linear stability analysis of the uniform state.
\bibitem{footnote1b}
{The magnitude and form of this term can be understood by imagining that the chemical is produced by each particle from a point source located at ${\mathbf r}_c-R{\mathbf p}$, where ${\mathbf r}_c$ 
is the colloid centre and $R$ its radius; this leads to $k_a\sim k_0R$ (compare also \cite{Saha2014}).}
\bibitem{footnote2}
Usually, the motion of self-phoretic colloids is driven by a slip velocity ${\mathbf v}_s$ which 
is proportional to a mobility, $M$, times the tangential chemical gradient $\nabla_{\parallel}c$. Chemotaxis (i.e., particle rotation set by $\nabla c$) then requires 
a non-zero integral $\int_S dS {\mathbf r}\times {\mathbf v}_s$ over the colloidal surface, $S$. This is generically satisfied by having (say) a non-uniform mobility over $S$, 
which is the default situation for a Janus particle.
\end{thebibliography}
\end{document}


\title{Supplemental Material to Clustering and Pattern Formation in Chemorepulsive Active Colloids}


\author{Benno Liebchen}\email[]{Benno.Liebchen@staffmail.ed.ac.uk}\affiliation{SUPA, School of Physics and Astronomy, University of Edinburgh, Edinburgh EH9 3FD, United Kingdom}
\author{Davide Marenduzzo}\affiliation{SUPA, School of Physics and Astronomy, University of Edinburgh, Edinburgh EH9 3FD, United Kingdom}
\author{Ignacio Pagonabarraga}\affiliation{Departament de F\'{i}sica Fonamental, Universitat de Barcelona-Carrer Mart\'{i} i Franqu\`{e}s 1, 08028-Barcelona, Spain}
\author {Michael E. Cates}\affiliation{DAMTP, Centre for Mathematical Sciences, University of Cambridge, Cambridge CB3 0WA, United Kingdom}

\maketitle


\subsection{Nondimensionalisation}
To understand 
the parameter regimes leading to the stationary and moving patterns observed in Figs.~2 and 3 of the main text, we perform a linear stability analysis of our model. 
We first reduce the parameter space by rewriting (1) of the main text in dimensionless quantities (denoted with tildes),
$\tilde t=k_d t;\;\; \tilde x= x\sqrt{k_d /D_\rho}$ and $\tilde \rho = \rho k_0 |\beta| v_0/(k^2_d D_\rho);\;
\tilde {\bf p}=v_0/(\sqrt{k_d D_\rho}){\bf p};\;\; \tilde c=v_0|\beta|/(k_d D_\rho)$. This leads to (now omitting tildes)
\ea
\dot \rho &=& -\nabla \cdot (\rho {\bf p}) + \nabla^2 \rho \nonumber \\ 
\dot {\bf p} &=& -\Gamma {\bf p} + \mathcal{D}_p \nabla^2 {\bf p} + s \nabla c - \Gamma_2 |{\bf p}|^2 {\bf p} \nonumber \\ 
\dot c &=& \mathcal{D}_c \nabla^2 c + \rho -c + \kappa \nabla \cdot (\rho {\bf p}). \label{eom3ru}
\eee
Here $\Gamma=\gamma/k_d;\; \mathcal{D}_p =D_{p}/D_\rho;\; \mathcal{D}_c=D_c/D_\rho; \; \kappa=k_a k_d/(k_0 v_0); \; \Gamma_2=\gamma_2 D_\rho/v_0^2$
are dimensionless control parameters and $\; s={\rm sign}(\beta) \label{ru}$ distinguishes positive ($s=1$) and negative ($s=-1$) chemotaxis.
The overall average colloid density, $\rho_0$, is a conserved variable and sets the initial reference state. 
Note that the patterning scenarios we report in the main text do not change qualitatively if this cubic saturation term is omitted and a nonlinear chemotactic coupling is instead used, 
e.g. $s\nabla c/|\nabla c|$, 
which prevents unlimited growth of $\nabla c$. We favor the cubic saturation term because then $\Gamma_2$ does not enter the linear stability analysis of the uniform state. 

As usual, the stability of the uniform state is determined by the temporal evolution of linearized perturbations around it of wavevector ${\bf q}$ 
quantified by their growth rates or dispersion relation, $\lambda({\bf q})$.
Linearising $(\rho,p,c)$ around the uniform solution $(\rho,p,c)=(\rho_0,0,\rho_0)$ of (\ref{eom3ru}) in one spatial dimension yields:
\1 
\left(\begin{matrix} \dot{\delta \rho} \\ \dot{\delta p} \\ \dot {\delta c} \end{matrix}\right)= 
M \left(\begin{matrix} \delta \rho \\ \delta p \\ \delta c \end{matrix}\right); \quad M=\left(\begin{matrix} \partial^2_x & -\rho_0 \partial_x & 0 \\ 0 & -\Gamma + \mathcal{D}_p \partial^2_x & s \partial_x \\ 1 & \kappa \rho_0 \partial_x &  \mathcal{D}_c \partial^2_x -1 \end{matrix}\right)
 \label{3dms}
\2
Inserting a plane wave Ansatz for all three fields (or employing a Fourier transformation) leads to the following
characteristic polynomial for $\lambda(z)$, with $z:=\I q$:
\ea P(\lambda)&:=&{\rm det}\left(M -\lambda \mathcal{I} \right) =(z^2-\lambda)(\mathcal{D}_c z^2 -\lambda -1)
 \nonumber \\ &\times & (\mathcal{D}_p z^2 -\lambda -\Gamma )
 + s \rho_0 z^2 \left( -\kappa \lambda + z^2 \kappa + 1\right) \label{cpol} \eee
As $P(\lambda)$ is a cubic polynomial, we can solve $P(\lambda)=0$ explicitly, leading to closed expressions for the dispersion relation $\lambda(z)$. However, 
the general result for $\lambda(z)$ is a complicated expression, whose physical meaning is not immediate.
\\Thus, we first evaluate the dispersion relations $\lambda(z)_{1,2,3}$ numerically which allows us to plot instability or phase diagrams (We show and discuss a typical example in the main text; see Fig.~4).
These diagrams typically show three different regimes representing (i) stability of the uniform state, (ii) growing stationary modes leading to stationary patterns 
and (iii) growing oscillatory modes leading to moving patterns. 
In the following sections we simplify our model to derive simple instability criteria allowing for a direct physical interpretation.

\subsection{Keller-Segel Model and Janus Instability}
To develop a quantitative understanding of the 
mechanism driving the growth of stationary clusters in Fig. 2e-h of the main text (and regime ii in Fig.~4) 
we now derive
a reduced model of (\ref{eom3ru}). 
We first assume that the active particle orientation, ${\bf p}$, relaxes faster than the conserved colloidal density and the chemical density,
whose spatial average approaches $\rho_0$ for long times $\langle c \rangle (t\rightarrow \infty) \rightarrow \rho_0 $, see (\ref{eom3ru}). Hence, we set $\dot {\bf p}\rightarrow 0$ 
and eliminate ${\bf p}$ adiabatically. 
Assuming further weak translational diffusion of 
${\bf p}$, ${\cal D}_p \ll 1$, 
and using that $|{\bf p}|^2 {\bf p} \approx {\bf 0}$ close to the uniform state,
we obtain ${\bf p}\approx (s/\Gamma) \nabla c$.
In this adiabatic regime the dynamics constitutes a generalized version of the Keller-Segel (GKS) model \cite{Keller1971,Theurkauff2012,Meyer2014}, as used to describe chemotactic interactions in microbial communities, 
extending it to account both for chemorepulsion (when $s=-1$), and anisotropic chemical production (when $\kappa\ne 0$). 
In terms of rescaled variables $\rho'\equiv \rho/\Gamma;\; c'\equiv c/\Gamma$, we can write (now omitting primes).

\ea
\dot \rho &=& -s \nabla \cdot (\rho \nabla c) + \nabla^2 \rho \nonumber \\ 
\dot c &=& s \kappa \nabla \cdot (\rho \nabla c) + \mathcal{D}_c \nabla^2 c + \rho -c \label{eom3r}.
\eee

We now perform a linear stability analysis of the GKS model. 
Linearising (\ref{eom3r}) around the uniform solution $(\rho,c)=(\rho_0,\rho_0)$
yields
\1 
\left(\begin{matrix} \dot{\delta \rho} \\ \dot {\delta c} \end{matrix}\right)= 
\left(\begin{matrix} 
\nabla^2 & -s \rho_0 \nabla^2  \\
1  & -1 + (D+ \kappa \rho_0 s) \nabla^2
\end{matrix}\right)
\left(\begin{matrix} \delta \rho \\ \delta c \end{matrix}\right) 
\2
Inserting a plane wave Ansatz and using $q:=|{\bf q}|$ leads to the dispersion relation 
\ea
&&2\lambda_\pm(q)=-1-q^2[1+\mathcal{D}_c + s \kappa \rho_0] \pm  \label{drel} \\
&&\sqrt{1+2q^2\left[\mathcal{D}_c + (2+ \kappa)\rho_0 s -1 \right] + q^4\left[\mathcal{D}_c + \kappa \rho_0 s -1 \right]^2} \nonumber
\eee
In the following we discuss the linear stability of the GKS separately for $s=1$ and $s=-1$ by direct analysis of (\ref{drel}).
An alternative and somewhat quicker way to derive most (but not all) of the following results is possible using the Routh-Hurwitz stability criterion \cite{Hurwitz1895,Willems1970}.

\begin{figure}
\begin{center}
\includegraphics[width=0.48\textwidth]{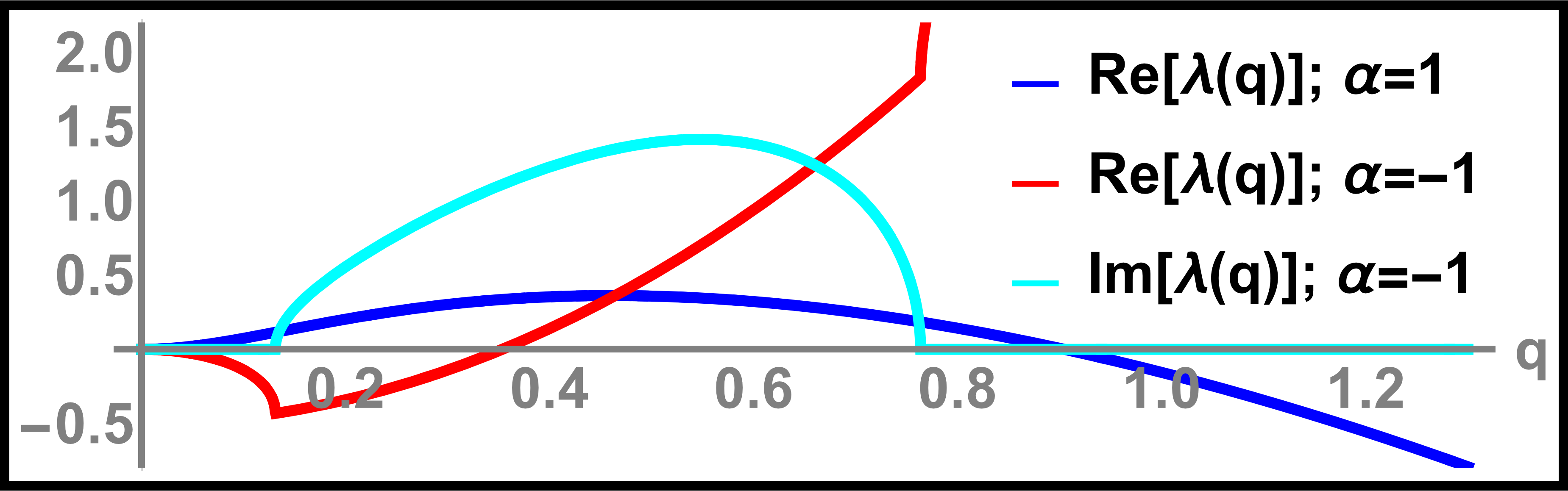} 
\caption{\small Dispersion relation for the generalized Keller-Segel model for $\mathcal{D}_c=\kappa=1; s=\alpha=\pm 1$.}
\label{fig1}
\end{center}
\end{figure}

\textbf{Chemoattraction:}
For chemoattraction or colloids turning towards high chemical concentration ($s=1$) the discriminant in (\ref{drel}) is always positive ($\mathcal{D}_c,\rho_0,\kappa>0$). Thus $\lambda(q)$ is real at all $q$ and we obtain the band of linearly unstable modes by solving $\lambda(q)=0$. This yields $q=\pm \sqrt{(\rho_0-1)/(\kappa \rho_0 + \mathcal{D}_c)}$ as the boundaries of the band of linearly unstable modes 
and additionally $q=0$ which is a double zero (We note that $\lambda(q\rightarrow \pm \infty) \rightarrow \infty$).
Hence, linearly unstable modes exist for $\rho_0 > 1$. In physical units, this instability criterion translates as $k_0 \beta v_0 \rho_0  > D_\rho k_d \gamma$; when this inequality is satisfied we observe phase separation in the GKS model (Eqs.~\ref{eom3r}).
The wave number of fastest growing mode follows from the conditions $\partial_q \lambda(q)=0$ and $\partial^2_q \lambda(q)<0$ as 
\1 q_{\rm max} = \pm \4{(1+b)\sqrt{\rho_0 b (\rho_0 + b -1)} - b(2\rho_0 + b -1)}{b(b-1)^2}\2
where $b=\kappa \rho_0 + \mathcal{D}_c$. This wave number determines the length scale of the droplets initially growing out of the uniform state before coarsening (similar to Fig.~2a, main text).

\textbf{Chemorepulsion:}
To determine the instability criterion for $s=-1$ we distinguish two cases. 
If the discriminant in (\ref{drel}) is positive, $\lambda_\pm(q)$ is real and the instability region follows from $\lambda_\pm(q)>0$. 
In contrast, if the discriminant is negative, the square root in (\ref{drel}) does not contribute to the real part of $\lambda_\pm(q)$ and 
the instability region is determined b $-1-q^2[1+\mathcal{D}_c - \kappa \rho_0]>0$. 
In this second case, ${\rm Im}(\lambda)\neq 0$ and the instability is oscillatory.

The discriminant vanishes at
\1 q^2_{\pm}=\left[a+\rho_0 \pm 2 \sqrt{\rho_0 a}\right]^{-1};\;\; a=1-\mathcal{D}_c +\rho_0(\kappa+1)\2 
and always has a minimum between $q_+$ and $q_-$ (except at $\mathcal{D}_c=1+\kappa \rho_0$). Thus, the discriminant is negative for $q^2\in [q^2_+,q^2_-]$.

We can now write the band of linearly unstable modes as
\ea && \{q| \lambda(q)>0; \; q^2\not \in [q^2_+,q^2_-] \} \cup \nonumber \\ && \{q| -1-q^2[1+\mathcal{D}_c - \kappa \rho_0]>0;\; q^2 \in [q^2_+,q^2_-] \} \eee
and rewrite it (after a lengthy but straightforward case by case analysis) in the following form:
\\$\bullet$ Regime 1: $\kappa < \mathcal{D}_c/\rho_0$: uniform state stable
\\$\bullet$ Regime 2: $\mathcal{D}_c/\rho_0 < \kappa \leq (1+\mathcal{D}_c)/\rho_0 + 1/\rho_0^2$: stationary wave instability with instability band $q^2 > (1+\rho_0)/(\kappa \rho_0-\mathcal{D}_c)$
\\$\bullet$ Regime 3: $ \kappa > (1+\mathcal{D}_c)/\rho_0 + 1/\rho_0^2$: oscillatory wave instability band $\{q| q^2 \in ([\kappa \rho_0 - (1+ \mathcal{D}_c)]^{-1},q^2_-) \}$ coexisting with a stationary instability band $\{q|q^2 \geq q_-^2\}$.
\\\\Thus, chemorepulsion leads to an instability of the uniform state if $\kappa > \mathcal{D}_c/\rho_0$ (in physical units this translates to: $k_a \beta \rho_0>\gamma D_c$).
This instability vanishes for $\kappa=0$; hence it requires anisotropic chemical production by the active colloids and we call it the 'Janus instability' (see main text). 
More specifically, the interplay of anisotropic chemical production and chemorepulsion generates an 
effectively negative chemical diffusivity ($-\kappa \rho_0 \nabla^2 c$ term in (\ref{eom3r})). If it dominates over the standard passive diffusion ($\mathcal{D}_c \nabla^2 c$), this initiates a spinodal instability. As summarized in the cartoon in Fig.~1 b (main text), the colloids, 
in turn, cross-diffuse towards minima in $c$ as a result of 
self-advection and chemorepulsion (the $\rho_0 \nabla^2 c$ term in (\ref{eom3r})). This cross-diffusion favours a colloidal density profile inverse to the chemical field (Fig.~2f,g in the main text). 

In contrast to the case of chemoattraction, for $s=-1$ the GKS model predicts a fastest growing mode occurring at zero wavelength ($\lambda(|{\bf q}|\rightarrow \infty) \rightarrow \infty$); 
suggesting the growth of very small clusters (compare Fig.~\ref{fig1}). 
In the full model (\ref{eom3ru}) however, rotational dynamics prevents this divergence and thus provides an essential feature for pattern formation
based on the Keller-Segel instability. 
This can be seen by performing an asymptotic expansion of (\ref{cpol}). Rewriting this equation in the form
$ P(\lambda)=c_0(z)+c_1(z)\lambda+c_2(z)\lambda^2 + c_3(z)\lambda^3 \label{opoly} $
and truncating all powers of $z$ but the highest one, separately for each coefficient $c_i(z)$ yields
\ea &-&\lambda^3 + z^2(1+\mathcal{D}_c + \mathcal{D}_p)\lambda^2 \nonumber \\ &+& z^4(-\mathcal{D}_p + \mathcal{D}_c (1+ \mathcal{D}_p) )\lambda + z^6 \mathcal{D}_c \mathcal{D}_p=0 \eee
Solving this equation leads to the following three dispersion relations for the three orthogonal modes which are linear combinations of the fields $\rho,p,c$ (these fields form a basis of the solution space to (\ref{3dms})) 
\1 \lambda_{1,2,3}(q \rightarrow \infty) = \left(-q^2,-\mathcal{D}_c q^2, -\mathcal{D}_p q^2 \right)\2
Accordingly, rotational dynamics suppresses the Janus instability at small wavelength and serves as an essential ingredient of pattern formation in our model. 
We note that the Janus instability is robust to sufficiently short-ranged repulsions. These interactions provide an alternative to rotational dynamics for
preventing the small wavelength collapse.
\\The orientational order we observed in Fig.~2f can be directly understood from (\ref{eom3ru}) as a 
consequence of colloidal particle number conservation:
(\ref{eom3ru}) imply that any current-free stationary state ($\dot \rho=0, {\bf j}_\rho=\rho {\bf p} - \nabla \rho=0$) yields ${\bf p}=\nabla \rho/\rho$, 
meaning that every local maximum (minimum) of $\rho$ corresponds to an aster (anti-aster) in the ${\bf p}$-field.

Besides unstable stationary modes the above linear stability analysis of the GKS model predicts also unstable oscillatory modes (case 3). 
It turns out, that these modes do not properly explain the moving patterns we observed in Fig.~3 of the main text. In particular, Fig.~4 of the main text shows that 
oscillatory instabilities may even occur for $s=-1,\kappa=0$, which cannot be explained based on the GKS model. 
Instead, the moving patterns we observed rely on a delayed response of the active colloids to changes in the chemical density. To see this 
quantitatively, we derive a minimal model for this 'delay-induced' instability in the next section.

\subsection{Delay-Induced Instability\label{dis}}
While the GKS model allowed us to understand the Janus instability mechanism leading to stationary patterns, we found above that a distinct instability mechanism must be responsible for
the moving patterns we observed in Fig.~3 of the main text. Specifically, in certain parameter regimes, 
a delayed response of the director field to changes in the chemical density leads 
to a qualitative failure of the 'adiabatic' $\dot {\bf p}\rightarrow 0$ approximation used to derive the GKS model.
This delay can cause an instability by itself as we discuss qualitatively in Fig.~1 of the main text. 
Quantitatively, this can be seen most
clearly,
when switching off rotational relaxation by setting $\Gamma =0$ (infinite delay). 
We also set $\kappa = 0$ in (\ref{eom3ru}) and for simplicity $\mathcal{D}_p=\mathcal{D}_c=0$ which yields the following minimal model which is, 
our simulations suggest, sufficient to generate delay-induced patterns 
of similar length scales to the full model:
\1 \dot \rho = -\nabla \cdot (\rho {\bf p}) + \nabla^2 \rho; \; \dot {\bf p}= -\nabla c + \Gamma_2 |{\bf p}|^2 {\bf p};\; \dot c = \rho -c \label{redmov} \2
A linear analysis of (\ref{redmov}) shows that, when $\beta<0$, an instability arises for $\rho_0 >1$; in physical units, this reads $k_0 |\beta| v_0 \rho_0> D_\rho k_d^2$. 
Accordingly, fast self-propulsion, strong chemotactic coupling, slow chemical decay and slow colloidal diffusion all promote the delay-induced instability.
This route to pattern formation requires self-propulsion and works for purely linear reaction dynamics, in crucial distinction to classical reaction-diffusion patterning
in biological and chemical systems~\cite{Turing1952,Murray2003}.

In the minimal model of (\ref{redmov}) with infinite delay time ($\Gamma=0$) the delay-induced instability first emerges at long wavelengths
while the fastest growing mode moves to shorter wavelength as $\rho_0$ increases.
We show now, that
reinstating a finite delay time via nonzero values of the angular relaxation rate $\Gamma$ generally
shifts the instability to short wavelengths, analogously to the situation described for the stationary Janus instability.
To this end, we reconsider the dispersion relation of the complete model (\ref{eom3ru}) and explore its long wavelength regime. 
The general behaviour of $\lambda(q)$ at small wavenumbers is nontrivial. For $\Gamma \ll 1$ (weak damping of long wavelength modes) however, 
the fastest growing mode occurs at small wavenumbers $q \ll 1, q \neq 0$ (which can be verified numerically) and we may
expand the general result for $\lambda(q)$ (the zeros of $P(\lambda)$ given by (\ref{cpol})) first in $\Gamma$ and then in $q$. 
In the special case $\kappa=0,s=-1$ (neglecting mixed terms of order $q^2 \Gamma$ and higher) this yields 
\ea
{\rm Re}[\lambda(q)]&=&-\Gamma/2 + q^2(\rho_0 -1 - \mathcal{D}_p)/2\nonumber \\ 
&& - q^4 \rho_0(2\rho_0-1 + 2\mathcal{D}_c - \mathcal{D}_p)/2 \label{rel3} 
\eee
for the real part of the the fastest growing mode in (\ref{3dms}).
Hence, the linear instability criterion reads
\1 (\rho_0 - 1 - \mathcal{D}_p)> 4\rho_0 \Gamma [2(\rho_0 + \mathcal{D}_c) - (1+ \mathcal{D}_p)] >0  \2
This shows that the delay-induced instability mechanism (explained in the main text) is robust with respect to finite 
diffusivity of the chemical and director field.
Note that the instability criterion reduces to requiring that 
$\rho_0 > 1$ when $\Gamma=\mathcal{D}_c = \mathcal{D}_p=0$ -- 
this is the exact instability criterion of the minimal model we just discussed.

Eq.~(\ref{rel3}) predicts the fastest growing mode as:
\1 q^2_{\rm max} = \4{\rho_0 - 1 - \mathcal{D}_p}{2\rho_0 [2\rho_0 + 2\mathcal{D}_c - \mathcal{D}_p -1]} \2
Far from the onset of instability ($\tilde \rho_0 \gg 1$) the wavelength of the fastest growing mode scales as $\rho_0^{1/2}$ (assuming $D_\rho \propto v_0^2$ according to ref. Theurkauff et al. in main text), 
or in physical units as $l^\ast \propto \sqrt{|\beta|v_0 k_0\rho_0/k_d^3}$. The selection of a finite wavelength for the patterns requires, therefore, a finite chemical decay rate.
In particular, the length scale of the fastest growing mode scales as $l \propto \sqrt{v_0}$ -- therefore the domain area of the corresponding two-dimensional clusters scales as $l^2 \propto v_0$ (as in
experiments \cite{Theurkauff2012,Buttinoni2013}).

The wavelength of the fastest growing mode can be computed more generally at the onset of instability 
(where ${\rm Re}[\lambda(q_{\rm max})]=0$ and we can determine this wavelength from the condition ${\rm Re}[c_0(z)]=0$ in (\ref{opoly})) 
and far away from the onset (where $\lambda \gg c_0$ and we can neglect $c_0(z)$ in (\ref{opoly})).
In general, the scaling for the length scale of the clusters, $l$, varies from $l \propto v_0^0$ (for $v_0\rightarrow 0$) to 
$l \propto v_0^2$ (for $v_0\rightarrow \infty$), which means that clusters based on chemorepulsion generically grow with increasing self-propulsion velocity. 

A low-order approximation of the imaginary part of the dispersion relation in $\Gamma$ and $q$ (neglecting $q \Gamma$ and higher order terms) leads to 
${\rm Im}\lambda(q)=\pm \sqrt{\rho_0}$. This shows that the frequency of the oscillation of long wavelength moving patterns is given (at lowest order) by 
$\omega = \sqrt{\rho_0}$. In physical units, this translates to $\omega \sim \sqrt{(|\beta| v_0 k_0 \rho_0)/D_\rho}$.
Hence, the frequency of the delay-induced oscillatory instability and the 
velocity of the travelling wave pattern (group velocity associated with the fastest growing mode) 
is, at leading order, independent of the self-propulsion velocity (assuming again $D_\rho \propto v_0^2$ 
and the wave number of the fastest growing mode $q_{\rm max}\propto 1/\sqrt{v_0}$).

\subsection{Amplitude Blinking}
In the main text, we briefly discussed that amplitude blinking of clusters (Videos 1,4 and 5) emerges as a consequence of a simultaneous growth of
stationary waves with a certain length scale $2\pi/q_s$ and oscillating waves with another length scale $2\pi/q_o$ and frequency $\omega$. 
Here, we discuss this in more detail and complement the discussion taking a physical viewpoint.

Amplitude blinking appears close to the transition between the Janus instability (regime (ii) in Fig.~4 (main text)) and the 
delay-induced instability (regime (iii)), where we have a discontinuous jump of the unstable-mode wavelength (color change in Fig.~4). 
The underlying dispersion relation (Fig.~4 (main text) inset) shows two distinct, coexisting regions of instability, an oscillatory branch at lower $q$ and non-oscillatory growth at higher $q$.
Both modes grow simultaneously and the fastest growing mode prevails, determining the wavelength of the emerging pattern. Near the transition, where 
the oscillatory perturbations grow only slightly slower than the fastest growing stationary mode,
we observe long-lived oscillations, visible as blinking in colloidal density patterns, 
on top of the Janus 
instability leading ultimately to stable colloidal clusters. 
The fact that in some cases the blinking goes on indefinitely suggests that it can be stabilised by nonlinear terms in the equations of motion, not captured in our linear stability analysis.

\begin{figure}
\begin{center}
\includegraphics[width=0.48\textwidth]{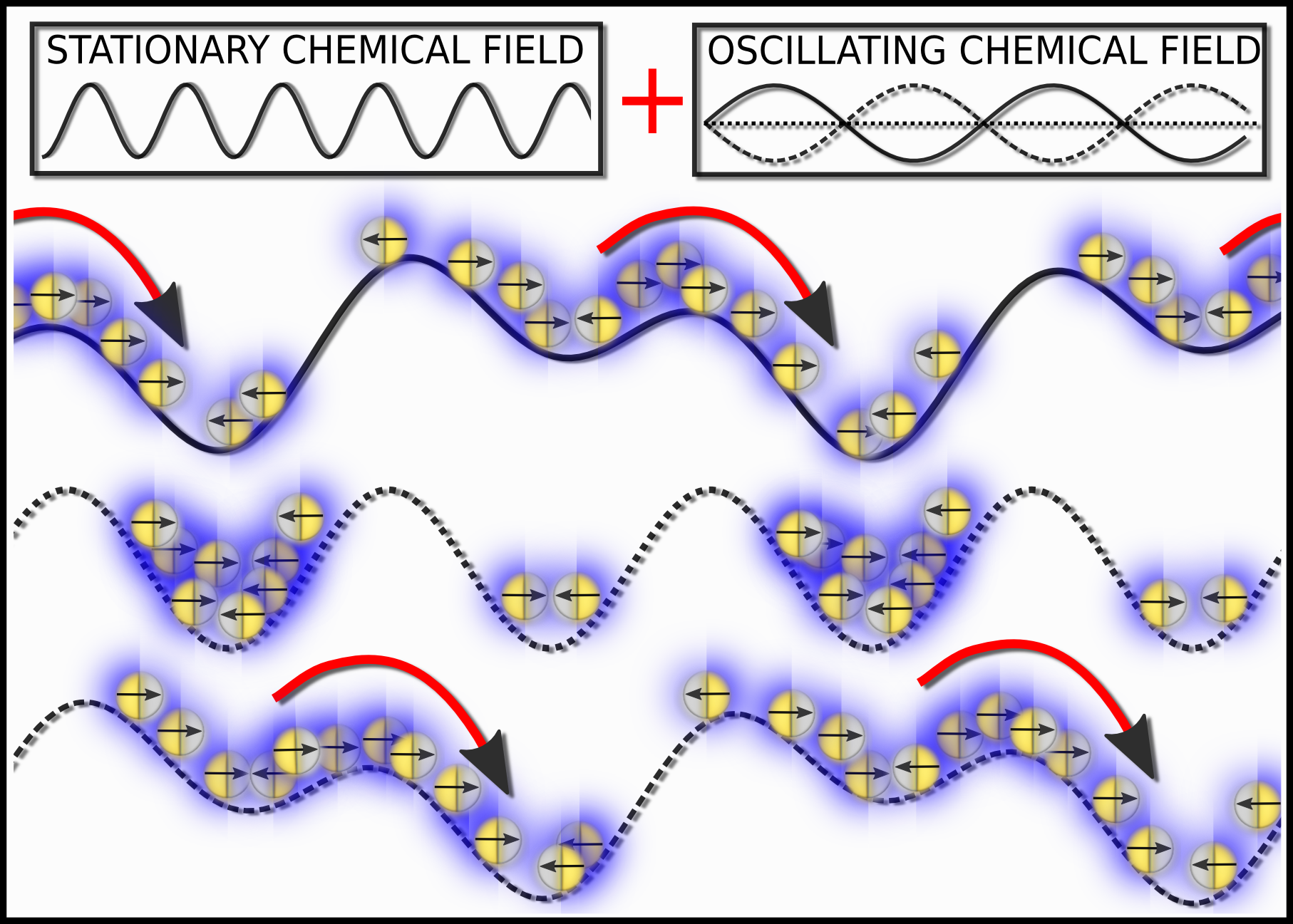} 
\caption{\small Cartoon of the physical mechanism leading to amplitude blinking of clusters.}
\label{figSI}
\end{center}
\end{figure}
In the linear regime, the superposition of the fastest growing modes of stationary and oscillatory plane wave fluctuations creates a chemical density field $c(x,t) \propto [\sin(q_s x) + \cos(\omega t) \sin(q_o x)$].
Fig.~\ref{figSI} shows such a field at three different times (black lines; for $q_o=q_s/2$). The local minima of this field correspond to the centres of clusters (asters) as created by the Janus instability. 
\\(i) In the upper panel (showing $c(x,t=0)$), chemorepulsive colloids move from clusters with comparatively high chemical density to adjacent ones with lower chemical density. 
As we see in the figure, this requires them to overcome (small) local density maxima of the chemical density field which is assisted by colloidal diffusion (favouring a uniform distribution of colloids). 
(ii) In the middle panel ($c(x,\pi/2)$), the minima of the chemical density field are equally deep. The corresponding colloidal clusters have different amplitudes (contain different numbers of colloids as a result of (i)) but 
the chemical density minima are deep enough to prevent a significant diffusive exchange of colloids.
(iii) As time evolves the chemical density increases at positions of the larger clusters (see lower panel, $c(x,\pi)$) 
which drives colloids again to adjacent clusters. This cycle repeats. Effectively, we observe a net particle exchange between different clusters resulting in the amplitude blinking we observed in Videos 1,4 and 5.

\subsection{Videos}
Videos 1-3: Parameters as in Figs.~2e-h, 3a-c and 3d (main text) respectively. 
\\Video 4: Parameters as in Fig.~3a-c but with no-flux (Neumann) boundary conditions instead of periodic boundary conditions. 
\\Video 5: Parameters as in Fig.~2e-h but with $\mathcal{D}_p=\mathcal{D}_c=1.104$ (This corresponds to the transition line between stationary and moving pattern 
where the growth rates of the fastest growing stationary and oscillating plane wave modes are identical.)